\documentclass[11pt,twoside]{article}

\usepackage{asp2006}
\usepackage{epsf}
\usepackage{psfig}
\usepackage{lscape}

\begin{document}

\title{Infrared Detection and Characterization of Debris Disks, Exozodiacal Dust, and Exoplanets:  The FKSI Mission Concept}   
\author{W. C. Danchi (1), R. K. Barry (1),  B. Lopez (2), S. Rinehart (1), O. Absil (5), J.-C. Augereau (3), H. Beust (3), X. Bonfils (3), P. Bord\'e (4), Denis Defr\`{e}re (5); Pierre Kern (3); P. Lawson (6); A. L\'eger (4), J.-L. Monin (3); D. Mourard (2); M. Ollivier (4), R. Petrov (2);  W. Traub (6);  S. Unwin (6); F. Vakili (2)}   
\affil{(1) NASA Goddard Space Flight Center, (2) Observatoire de la C\^ote d'Azur (OCA), (3) Laboratoire d'Astrophysique de Grenoble (LAOG), 
(4) Institut d'Astrophysique Spatiale, Orsay (IAS), (5) Universit\'e de Li\'ege and the FKSI Consortium, (6) Jet Propulsion Laboratory, California Institute of Technology.
}    
\begin{abstract}
The Fourier-Kelvin Stellar Interferometer (FKSI) is a mission concept for a
nulling interferometer for the near-to-mid-infrared spectral region. FKSI is
conceived as a mid-sized strategic or Probe class mission. FKSI has been
endorsed by the Exoplanet Community Forum 2008 as such a mission and
has been costed to be within the expected budget.
The current design of FKSI is a two-element nulling interferometer. The
two telescopes, separated by 12.5m, are precisely pointed (by small
steering mirrors) on the target star. The two path lengths are accurately
controlled to be the same to within a few nanometers. A phase shifter/beam combiner (Mach-Zehnder interferometer) produces an output beam consisting of the nulled sum of the target planet's light and the host star's light. When properly oriented, the starlight is nulled by a factor of 10$^{-4}$, and the planet light is
undiminished. Accurate modeling of the signal is used to subtract the
residual starlight, permitting the detection of planets much fainter than the
host star.  The current version of FKSI with 0.5-m apertures and waveband 3-8 $\mu$m
has the following main capabilities: (1) detect exozodiacal emission levels to 
that of our own solar system (1 Solar System Zodi) around nearby F, G, and K, stars; 
(2) characterize spectroscopically the atmospheres of a large number of known non-transiting  planets; 
(3) survey and characterize nearby stars for planets down to 2 R$_{\earth}$ from just inside the habitable zone and inward.
An enhanced version of FKSI with 1-m apertures separated by 20 m and cooled to 40 K, with science waveband 5-15 $\mu$m, allows for the detection and characterization of  2 R$_{\earth}$  super-Earths and smaller planets in the habitable zone around stars within
about 30 pc.
\end{abstract}
%
\section{Introduction}
A great deal of progress has been made in the study of exoplanetary systems in the last decade, with more than 400 planets discovered.  Some key questions of immediate importance are: (1) What are the atmospheres of non-transiting extrasolar planets composed of, and what are the physical conditions in their atmospheres?  (2) How many 2-R$_{\earth}$ planets are in our solar neighborhood and what are their atmospheres composed of? (3) What is the amount of exozodiacal dust around nearby stars?  Is it comparable to that of our own Solar system? (4) What is the spatial distribution of debris material around stars like our sun, how is it sculpted by the presence of planets, and how does this material form and evolve?  The thermal infrared spectral region has distinct advantages compared to the visible region because the star-to-planet contrast ratio in the habitable zone is about 3 orders of magnitude less than at visible wavelengths, and the thermal emission from warm (room temperature) material emits most strongly in the infrared.
\begin{figure}[!ht]
\begin{center}
 \resizebox{5cm}{!}{\includegraphics  {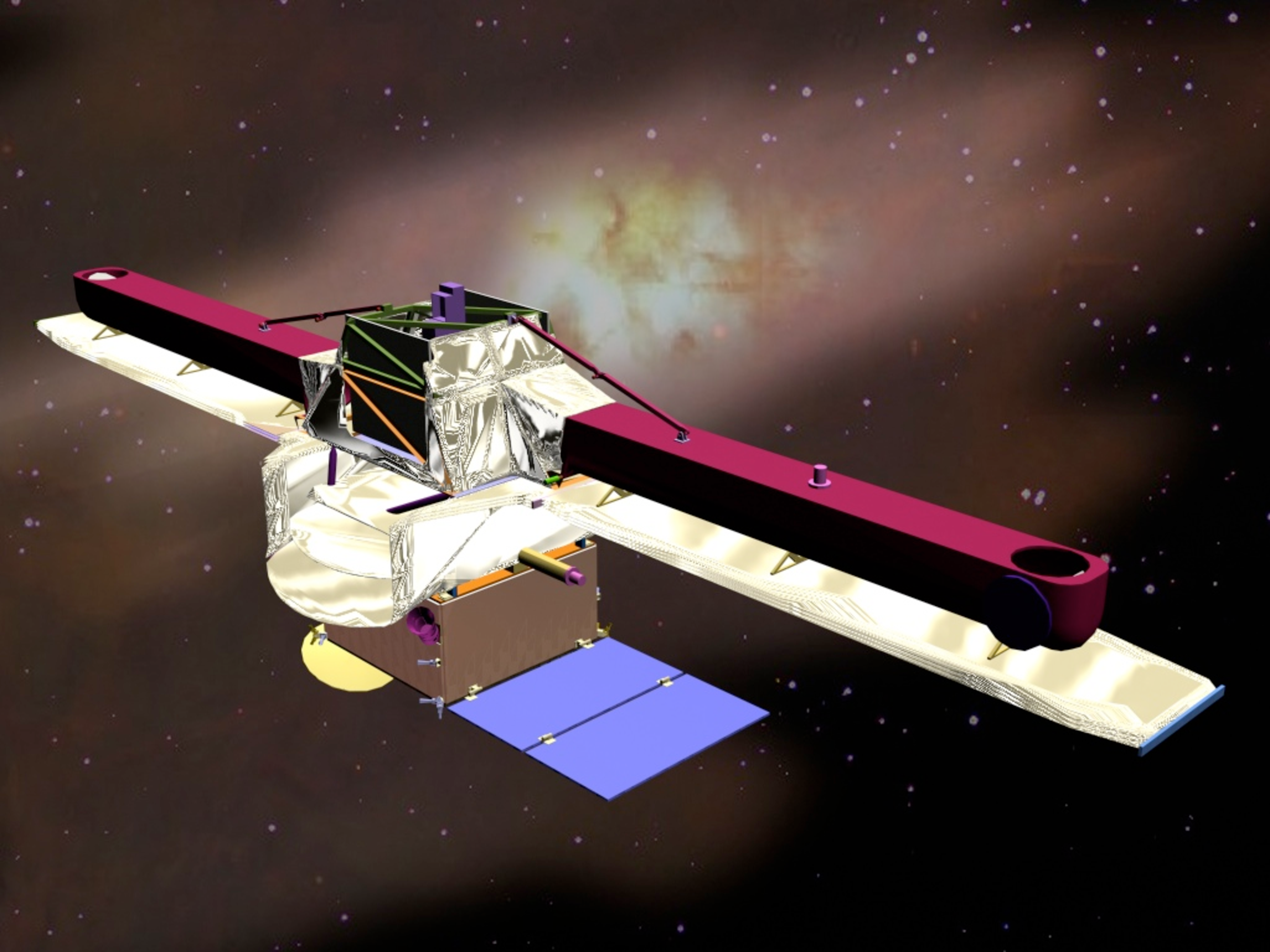}}
\caption{ Artists conception of current FKSI system as in operation at L2.}
\end{center}
\label{fig1}
\end{figure}
The Fourier-Kelvin Stellar Interferometer (FKSI) is a structurally connected infrared space interferometer with 0.5 m diameter telescopes on a 12.5-m baseline, and is passively cooled to 60 K.  FKSI operates in the thermal infrared from 3-8 $\mu$m in a nulling (or starlight suppressing) mode for the detection and characterization of faint material around relatively bright stars such as  exoplanets, debris disks, emission levels of  extrasolar zodiacal dust disks.  Figure 1 displays an artist's conception of the spacecraft in operation at L2.  FKSI will have the highest angular resolution of any infrared space instrument ever made with its nominal resolution of 40 mas at a 5 $\mu$m center wavelength.  This resolution exceeds that of Spitzer by a factor of 38 and JWST by a factor of 5.  The FKSI mission is conceived as a Òprobe classÓ or Òmid-sizedÓ strategic mission that utilizes technology advances from flagship projects like JWST, SIM, Spitzer, and the technology programs of TPF-I/Darwin.  
%
\section{History and Current Status}
The FKSI mission concept has been under development for a number of years at NASA's Goddard Space Flight Center (Danchi et al. 2003, 2007) and it has a budget that fits into the strategic mission or Probe-class category with a lifecycle cost of around \$600-800 million US dollars.   During the last few years technology development funded by NASA and ESA for TPF-I, Darwin, and JWST have retired most of the major risks.  Most of the key technologies are at a Technical Readiness Level (TRL) of 6 or greater, which means that, if funded, FKSI could enter into Phase A within the next two years (Danchi et al. 2008). The FKSI mission is designed to answer major scientific questions on the pathway to the discovery and characterization of Earth-twins around nearby F, G, and K main sequence stars.  

The FKSI mission is realistic, cost-effective, and low risk.  First, it has been studied extensively since 2002 and its baseline design is at an advanced state of development.  Consequently a detailed mass chart exists, which has been used in connection with both grass roots and parametric cost estimates.  Second, it is cost effective because many of its components can be adapted from or simply copied from those of JWST.  These include sun shades, cryocoolers, detectors, mirrors, and precision cryogenic structures. Indeed these components are at Technical Readiness Level (TRL) 6 since JWST passed its Technical Non-Advocate Review (TNAR) and is in Phase C/D.  Third, the only significant components that are below TRL 6 are optical fibers used for wavefront cleanup and the nuller instrument itself both of which require cryogenic testing. These final technology activities can be accomplished during Phase A.

The FKSI mission  has broad support in the US and European communities and has been endorsed by the Exoplanet Community Forum as a Òprobe classÓ or Òmid-sizedÓ strategic mission by the US community.  In France, FKSI has been included in the plans by the French space agency CNES as a Òmission of opportunity.Ó  It has also been included in plans by the European Space Agency (ESA) in a similar context. 
\begin{figure}[!ht]
\begin{center}
 \resizebox{6cm}{!}{\includegraphics  {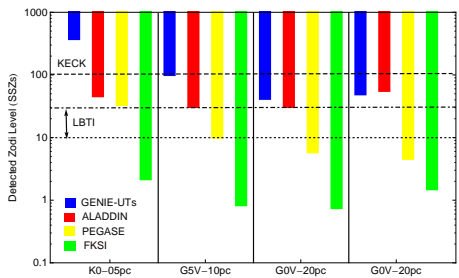}}
\caption{ Comparison of exozodiacal detection limits for two ground-based concepts (GENIE with UTs on the VLTI and ALADDIN), and two space mission concepts (PEGASE and FKSI) (adapted from Defr\`{e}re et al. 2008, A\&A, 490, 435).}
\end{center}
\label{fig2}
\end{figure}
\begin{figure}[!ht]
\begin{center}
\resizebox{6cm}{!}{\includegraphics{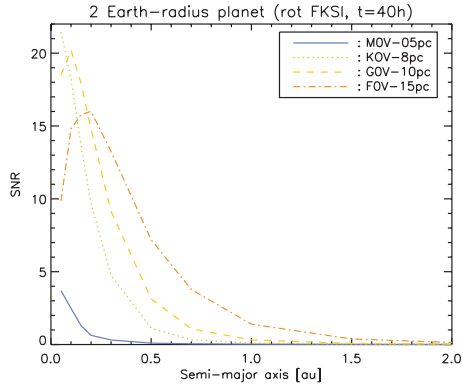}}
\caption{ Simulations for 2 R$_{\earth}$ super-Earth detection show that FKSI can detect many rocky planets around nearby F, G, K, and M stars (Defr\`{e}re et al. 2009, private communication).}
\label{fig3}
\end{center}
\end{figure}
\section{Performance Estimates for the Current FKSI Design}
One of the most vexing questions in the study of exoplanetary stellar systems is the location and amount of emission of 
warm dust in the habitable zone of these stars, analogous to the zodiacal light in our own Solar System.   This emission around nearby stars is called exozodi emission and is measured in units relative to that of our own Solar System (defined as one Solar System Zodi, or SSZ).  
Current consensus in the exoplanet community is that a small (non-flagship) space based mission is required to measure exozodi levels to that required to properly scope and specify a flagship mission (see for example the panel summary on this topic in this volume by  Absil et al.)
Figure 2 is a comparison of the ability of ground- and space- based concepts to measure this emission for nearby solar type stars.  This figure demonstrates that FKSI is the only instrument capable of measuring exozodi levels down to one SSZ in the habitable zone, and given the relatively short integration times required, within a few months of operation FKSI can measure the zodi levels for all TPF and Darwin target stars of interest. 

Another major scientific area is characterizing the atmospheres of exoplanets discovered through radial velocity and other techniques.  Currently only a small fraction of exoplanet atmospheres can be studied spectroscopically using transit methods, and these exoplanets are largely the ones with short periods that are very close to their stars.  Not being limited to transiting planets, FKSI  will greatly extend the size of the sample, which will have an enormous  impact on models of planet formation and evolution. A third area is the search for rocky planets of size  2-R$_{\earth}$ super-Earths or smaller in the habitable zone of nearby stars.  Figure 3 displays a simulation of the FKSI's capability for super-Earth detection.
\begin{figure}[!ht]
\begin{center}
\resizebox{9cm}{!}{\includegraphics{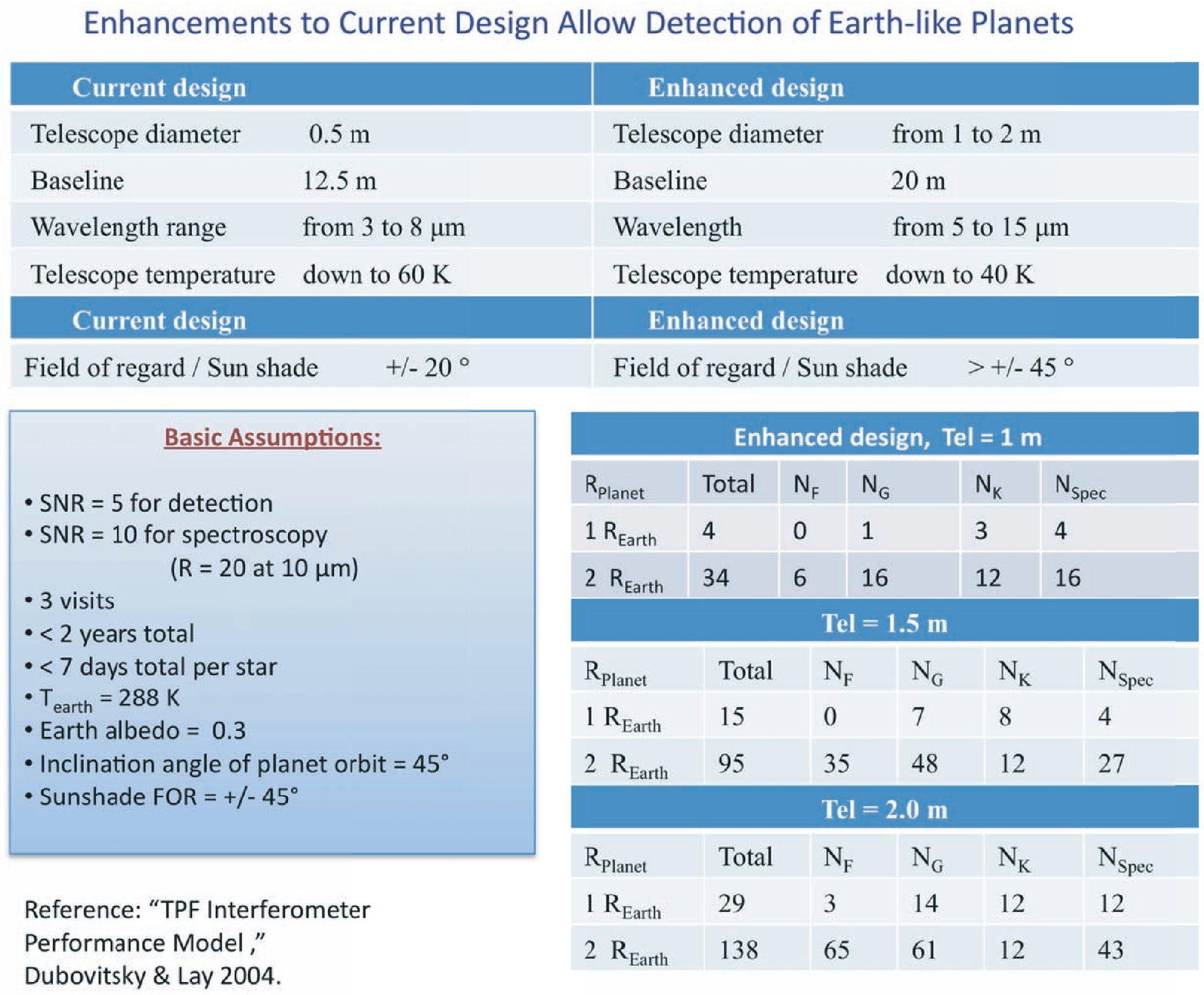}}
\caption{Simulations of FKSI performance with  1-2 m class telescopes at 40K and a 20-m baseline demonstrate that many 2 R$_{\earth}$ super-Earths and a few Earth-twins can be discovered and characterized within 30 pc of the Sun.}
\label{fig4}
\end{center}
\end{figure}
\begin{figure}[!ht]
\begin{center}
\resizebox{6cm}{!}{\includegraphics  {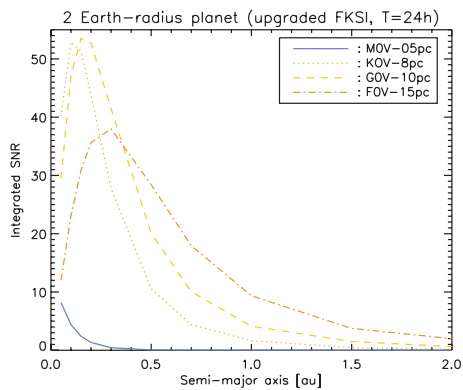}}
\caption{Simulations for 2 R$_{\earth}$ super-Earth detection for the enhanced FKSI (1-m apertures, 40 K telescope temperature, 20-m separation) show that FKSI can detect many  rocky planets {\em in the habitable zone} around nearby F, G, K, and M stars (Defr\`{e}re et al. 2009, private communication). }
\label{fig5}
\end{center}
\end{figure}
\section{Performance Estimates for an Enhanced Design}
As part of our submission to the Astro2010 Request for Information for the Program Prioritization Panel (PPP) on Electromagnetic Observations from Space (EOS), we began to look at an enhanced version of FKSI, with parameters more ideally suited for exoplanet characterization, namely moving the wavelength band from 3-8 $\mu$m to 5-15 $\mu$m, thus being centered on 10 $\mu$m,  near the emission peak of the Earth's spectrum.  

We used the TPF performance simulator of Dubovitsky and Lay (2004) and computed the number of 1-R$_{\earth}$ and 2-R$_{\earth}$ sized planets we could detect using standard assumptions that were used for similar studies of the flagship TPF Interferometer mission, as a function of aperture size from 1-m to 2-m diameter, with a 40 K telescope temperature, 20-m separation, and $\pm$ 45$\deg$ field-of-regard.  The key result from these simulations is that even with a 1-m aperture diameter, we were able to detect and characterize up to four 1-R$_{\earth}$ Earth-twins  around very nearby G and K stars, and detect 34 2-R$_{\earth}$ super-Earths and characterize 16 of them.  Figure 5 displays a summary comparison of key design features of the current design compared to those of the enhanced design, as well as the assumptions, and the key results of the simulation.  Figure 5 displays the results of a more complex numerical simulation of the performance of FKSI for super-Earth detection (see Defr\`{e}re et al. 2008 for more details of the simulation), which is consistent with the results presented in Figure 4.
\begin{figure}[!ht]
\begin{center}
        \resizebox{9cm}{!}{\includegraphics  {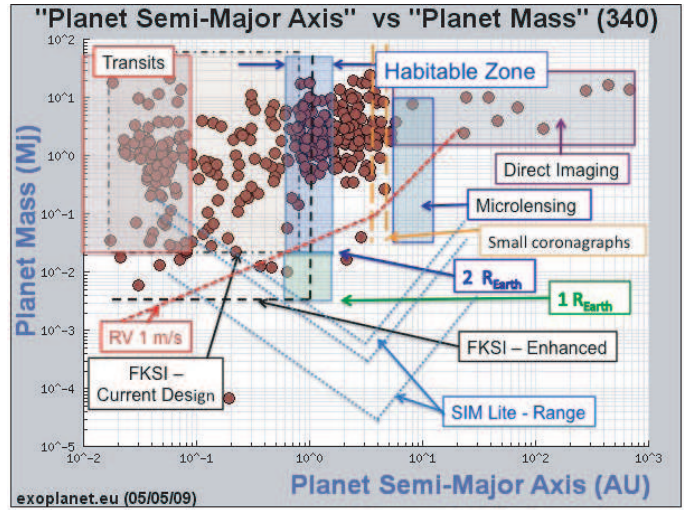}}
\caption{ Discovery space for exoplanets for FKSI and other mission concepts and techniques. }
\label{fig6}
\end{center}
\end{figure}
Figure 6 displays the discovery space of FKSI for exoplanets as a function of semi-major axis and compares it to other missions.   
The enhanced version of the FKSI system, with its approximate discovery space marked by the dark dashed lines, has a greatly enhanced phase space for detection compared to the current design, and is the only mission within the projected Probe cost cap that is capable of detecting Earth-twins in the habitable zone of nearby stars.  Figure 7 displays an engineering model of the enhanced design of FKSI, which has been shown to be technically feasible and within the projected cost envelope of the Probe class mission category.
\begin{figure}[!ht]
\begin{center}
\resizebox{6.5cm}{!}{\includegraphics{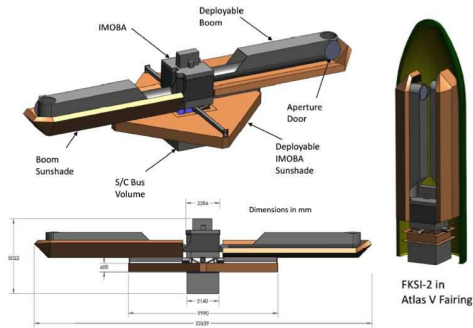}}
\caption{Engineering model of the enhanced FKSI design, called FKSI-2.}
\label{fig7}
\end{center}
\end{figure}
%
%
%
%
%
%


\begin{thebibliography}{}
\bibitem{Danchi03}Danchi, W. C., et al., 2003, ApJ, 597L, 570.
\bibitem{Danchi07}Danchi, W. C., \& Lopez, B.  2007, Compte Rendu Physique, 8, 396.
\bibitem{Danchi08}Danchi, W. C., et al. 2008, Proc. SPIE, vol. 7013.
\bibitem{Def08} Defr\`{e}re, D., et al., A\&A, 490,  435.
\bibitem{Dub04}Dubovitsky, S. \& Lay, O., Proc. SPIE, 5491, 284. 
\end{thebibliography}
\end{document}